\documentclass[graybox]{svmult}

\usepackage{mathptmx}       
\usepackage{helvet}         
\usepackage{courier}        
\usepackage{type1cm}        
\usepackage{makeidx}         
\usepackage{graphicx}        
\usepackage{multicol}        
\usepackage[bottom]{footmisc}

\makeindex             

\begin{document}

\title*{Study of a model for the distribution of wealth}
\author{Yves Pomeau and Ricardo L\'opez-Ruiz}
\institute{Yves Pomeau \at Department of Mathematics, University of Arizona, Tucson, USA \email{pomeau@lps.ens.fr} 
\and Ricardo L\'opez-Ruiz \at Department of Computer Science \& BIFI, University of Zaragoza, Zaragoza, Spain \email{rilopez@unizar.es}
}
\maketitle

\abstract*{An equation for the evolution of the distribution of wealth in a population of economic agents making binary transactions with a constant total amount of ``money" has recently been proposed by one of us (RLR). This equation takes the form of an iterated nonlinear map of the  distribution of wealth. The equilibrium distribution is known and takes a rather simple form. If this distribution is such that, at some time, the higher momenta of the distribution exist, one can find exactly their law of evolution. A seemingly simple extension of the laws of exchange yields also explicit iteration formulae for the higher momenta, but with a major difference with the original ite-ration because high order momenta grow indefinitely. This provides a quantitative model where the spreading of wealth, namely the difference between the rich and the poor, tends to increase with time.}

\abstract{An equation for the evolution of the distribution of wealth in a population of economic agents making binary transactions with a constant total amount of ``money" has recently been proposed by one of us (RLR). This equation takes the form of an iterated nonlinear map of the  distribution of wealth. The equilibrium distribution is known and takes a rather simple form. If this distribution is such that, at some time, the higher momenta of the distribution exist, one can find exactly their law of evolution. A seemingly simple extension of the laws of exchange yields also explicit iteration formulae for the higher momenta, but with a major difference with the original iteration because high order momenta grow indefinitely. This provides a quantitative model where the spreading of wealth, namely the difference between the rich and the poor, tends to increase with time.}

\section{Introduction}

This communication follows the Noma-13 conference in September 2013, an enjoyable and fruitful meeting where one of us (YP) had a chance to hear of the model considered below \cite{zareq}. This model describes the evolution of the distribution of wealth in a population of individuals doing business pairwise. After each exchange there is a redistribution of money between the two individuals, without total loss or gain. A feature of this model, the ``Z-model" (with Z for Zaragoza) is its simple equilibrium solution (written below). Under its law of evolution, this equilibrium solution is stable and so attract most, if not all initial conditions satisfying convergence conditions (finite total probability and finite total wealth) \cite{zareq.1}. Moreover, an $H$-theorem is valid for this model \cite{apenko2013}. We show below that the evolution of higher momenta (mean square value, mean cubic value, etc.) of the wealth can be computed exactly, obviously under the condition that those momenta exist. We consider also situations where the momenta do not converge beyond a given order. An anonymous referee pointed out that something called ``q-model" has equations similar to the Z-model. Those q-models aim at describing the distribution of stress in random set of solid grains in contact with neighbours in such a way that the downward push of the weight of a grain and of the grain above it is distributed more or less randomly between its neighbours underneath. In this theory the equivalent of the time of the Z-model is played by the vertical direction and the time-iteration amounts to move down the pile to find the distribution of stress on grains. Even though the equations of this q-model look like the ones of the Z-model, their physical meaning is quite different. The interested reader may get a list of papers on the subject in the reference list of the lecture notes published in \cite{bouchaud}.  Moreover the q-model, in order to get a row to row equation of iteration like the one of the Z-model has to assume that the vertical force on beads on the same horizontal row are statistically independent, which is presumably needed to get at the end something like a hyperbolic system, although the Cauchy-Poisson equations for regular elasticity are elliptic.  

Because of its simple mathematical structure it makes sense to extend the Z-model by keeping the possibility of an exact solution for the momenta. This can be done with a straightforward extension maintaining the basic properties of conservation of the total probability and the total wealth. Although this modified Z-model looks very much like the original and reduces to it continuously as a parameter changes, it has completely different properties. In particular it shows an increase of the fluctuations of wealth as time goes, a rather unexpected property,  absent in the original model. This makes the matter of Section \ref{sec:genZ}.  In this respect the inequality of wealth, as studied below makes only a small part of this big subject, but it is at least one that one can try to describe quantitatively.

Motivated by this consideration of momenta, we look in Section \ref{sec:divmom} at what happens in the Z-model when the momenta do not converge, specifically when the distribution of wealth decays algebraically for large values so that momenta do not exist, at least initially, beyond a certain power (This might be related to what is called Pareto law, Pareto \cite{pareto} having predicted that the natural distribution of wealth decays algebraically for large values, a property of the mZ-model studied below). An interesting result of this analysis is that, after a certain number of iterations (namely after a finite amount of time) higher momenta converge although they diverged initially. Somehow, without venturing into the area of political science, this looks like the exact opposite of what is predicted sometimes (without relying on objective modelisation as much we can tell): fewer and fewer individuals get richer and richer although the other ones get poorer and poorer as time goes. This could have other explanations of course, like what is called the redistribution of wealth by the tax system in modern economies.

 We shall explain first how to solve ``exactly" the moment problem, for a probability distribution decaying fast enough at infinity and then look at what happens if, initially, this probability distribution decays algebraically for large values. 
 
 In section \ref{richest} we give the probability distribution of the wealth of the ``richest man", namely the largest wealth of a given finite number of  agents with a given probability distribution of the wealth with agents taken at random in the population.  An explicit expression of this probability distribution of the maximum of wealth, with its limit in the case of a large number of agents, is given.
 
 The last Section is a Summary and Conclusion section.

\section{The Z-model} 
\label{sec:Zmodel}

In this model one considers a positive variable, with various names, $x$, $u$, etc, is for the amount of money owned by an individual. This amount changes in the course of time because of random exchanges between the individuals taking place at discrete time, in a  synchronous way in the system. The fundamental quantity is $p_t(x)$, the probability that an individual taken at random in the population has an amount $x$ at time $t$. At the next time step $(t +1)$, due to the binary exchanges, $p_t(x)$ has changed according to the law of iteration found of reference \cite{zareq}: 

\begin{equation} 
p_{t+1} (x) = \int \int_{S(x)} {\mathrm{d}}u {\mathrm{d}}v \frac{p_{t} (u) p_{t} (v)}{u + v}
\mathrm{,}
\label{eq:iter}
\end{equation}
The domain of integration in Equation (\ref{eq:iter}) is defined by  
$$ S(x) = \{ (u, v), u, v>0, u + v > x \}\mathrm{.}$$
This integral equation is for a function of $x$, positive variable. Because $p(.)$ is a probability distribution, it has to be positive or zero. Moreover it is normalised in such a way that $ \int_0^{\infty} {\mathrm{d}}u p_t (u) = 1$,  and $t$ is a discrete index representing time.  
This law of evolution of the wealth is derived as follows. Suppose two individuals, each one with the same probability of wealth, say $p(u)$, put their money in the same basket. Then the probability distribution for what is in the basket (the amount $w$) is 
$$q(w) = \int_0^{\infty}  {\mathrm{d}}v p(v) p(w - v) H(w-v) \mathrm{,}$$
where $H(.)$ is Heaviside function, zero for a negative argument and one otherwise. Suppose we share between two individuals the amount $w$ by taking randomly a value in $[0, w]$, give it to the first individual and the rest to the other. The probability distribution of what is taken by anyone of those individuals is 
$$ r(s) = \frac{\chi_{w}(s)}{w} \mathrm{,}$$
where $\chi_{w}(s)$ is the characteristic function of the interval $[0, w]$. By extending this simple formula to the probability distribution $q(w)$ of the values of $w$, as derived above, one obtains:
$$ r(s) = \int_0^{\infty} \frac{ {\mathrm{d}}w}{w} H(w - s)  \int_0^{\infty}  {\mathrm{d}}v p(v) p(w - v) H(w-v)   \mathrm{,}$$
After rearranging the integrals one finds
$$ r(s) = \int_{0}^{\infty}{\mathrm{d}} v' \int_{s-v'> 0}^{\infty} p(v') p(u')  \frac{{\mathrm{d}} u'}{u' + v'} \mathrm{,}$$
which is a form of the right-hand side of Equation (\ref{eq:iter}).

Equation (\ref{eq:iter}) can be integrated explicitly, at least in some sense. Let us define the moments of $p_t(x)$ as
\begin{equation} 
m_k (t) = \int {\mathrm{d}}u u^k p_t (u) 
\mathrm{.}
\label{eq:momendef}
\end{equation}
We consider first the case where all momenta converge. In Section  \ref{sec:divmom} we discuss the situation where some momenta do not exist at a given time because the integral (\ref{eq:momendef}) diverges at $k$ large, which is well possible because the ``physical" constraints on  $p(u)$ is to have well defined (not diverging) values of $m_0$ and $m_1$ only. 
From Equation (\ref{eq:iter}) one derives the following equation for the momenta of $p_{t+1} (.)$ as a function of the momenta of $p_{t} (.)$: 
\begin{equation} 
m_k (t+1) = \frac{1}{k + 1} \Sigma_{0\leq l \leq k} C_k^l m_{k-l} (t) m_{l} (t) 
\mathrm{,}
\label{eq:momenteq}
\end{equation}
where $C_k^l = \frac{ k!}{(k - l)! l!}$ are the binomial coefficients. 
This shows that the momenta of order $k$ at time $(t + 1)$ can be found if the momenta of smaller power at time $t$ are known. The formula is also consistent with the fact that $m_0 = 1$ at any time and that $m_1$ is a conserved positive constant (called later $m_1$). Let us look at the equation for $m_2$. It reads:
\begin{equation} 
m_2 (t+1) = \frac{2}{3} ( m_2(t) + m_1^2)
\mathrm{,}
\label{eq:momenteq2}
\end{equation}
Because this equation is linear with respect to $m_2$ it can be integrated at once with the result (supposing $m_2 (0)$ given):
\begin{equation} 
m_2 (t) = \left(\frac{2}{3}\right)^t  m_2(0) + 2 m_1^2  \left[ 1 -  \left(\frac{2}{3}\right)^t \right] =  \left(\frac{2}{3}\right)^t (m_2(0) - 2 m_1^2) + 2 m_1^2
\mathrm{,}
\label{eq:momenteq2sol}
\end{equation}
The higher momenta can be computed also explicitely as functions of the initial data for the lower order momenta, the result becoming more and more cumbersome as the order increases. At third order one has:
\begin{equation} 
m_3 (t+1) = \frac{1}{2} ( m_3(t) + 3 m_2(t) m_1)
\mathrm{,}
\label{eq:momenteq3}
\end{equation}
Let 
$$S_3(t) =  \frac{3}{2} m_2(t) m_1 \mathrm{.} $$ Therefore $$m_3(t) =  \left(\frac{1}{2}\right)^t  \left[ m_3(0) + \Sigma_{0\leq \theta \leq t} 2^{\theta} S_3 (\theta - 1)\right] \mathrm{,} $$ is a solution for $m_3(t)$ as a function of $m_1$, $m_2(0)$ and $m_3(0)$. The sums can be done explicitly because they involve geometric series. 
 The method of integration just explained does not work if one takes momenta with noninteger exponents because there is no finite equivalent of the binomial formula for such noninteger power.

 \section{Definition and solution of a generalized Z-model}
 \label{sec:genZ}
 
 The Z-model can be generalized in the following way. In the original formulation, each of the two partners in a transaction have a random amount $u$ and $v$. During the transaction they put first the whole amount $(u+v)$ in a basket and then share its content randomly. The Z-model describing this satisfies the constraint that the total probability is one and that the total money is also conserved. This model has also the property that the equilibrium solution (namely the distribution of wealth such that $p_t(u) = p_{t+1}(u)$) is known explicitly and is 
 $$p_{eq} (u) = \frac{1}{ m_1} e^{-\frac{u}{m_1}} \mathrm{,} $$
 
 Below we suggest a modified recursion relation, analogous to the one given in Equation (\ref{eq:iter}) but such that no simple expression of the equilibrium distribution can be found, even though the mass and first momentum $m_1$ is conserved (we keep the same notation, $m_k(t)$ for the $k$-th moment in the mZ-model, defined below, as in the Z-model). This model reads: 
 \begin{equation} 
P_{t+1} (x) = \int \int_{S_a(x)} {\mathrm{d}}u {\mathrm{d}}v \frac{P_{t} (u) P_{t} (v)}{a u + (2-a) v}
\mathrm{,}
\label{eq:iterm}
\end{equation}
In this equation, $a$ is a real parameter, between $0$ and $2$, and $S_a(x)$ is defined by the condition $x < a u + (2-a) v$.  In this model at the time of the transaction between the two individuals, one of the individual puts $(au)$ in the basket (instead of $u$ in the Z-model) and the other puts $(2-a)v$ in the basket, instead of $v$. Although this model is apparently not conservative, this is not the case. If we consider the symmetrical interaction for the pair of agents $(v,u)$, in this case the first agent will put $(av)$ in the basket and the second one $(2-a)u$. For both trades, those of the pairs $(u,v)$ and $(v,u)$, the total money to share in the basket is $2(u+v)$, then the total wealth is conserved. It can be interpreted that the excess of money in one of the trades is injected to cover the lack of money in the other trade. This is just one of the functions done by the bank system. Therefore, perhaps this is not such an unrealistic model because, nowadays (and very likely before), banks and even States rent money they do not really have and do that within  constraints based on multiplicative factors of their actual wealth. 

Like the Z-model, the modified Z-model (or mZ-model) defined by the iteration (\ref{eq:iterm}) satisfies the  constraints of conservation of $m_0$ and $m_1$ if $m_0 = 1$. From simple algebra, one finds:
$$ m_0(t+1) = m_0(t)^2 \mathrm{,} $$ and 
$$ m_1(t+1) = m_0(t) m_1(t) \mathrm{.} $$
Therefore the first two momenta are constant if $m_0 = 1$ and if $m_1$ converge, as we assume it. 
Contrary to the case of the Z-model, there is no simple equilibrium solution. However it is possible to derive many properties of this equilibrium from the equations for the moments. This is because the denominator in the iteration formula is a linear function of $u$ and $v$ like in the Z-model. The recursion relation for the second moment is:
 \begin{equation} 
m_{2} (t + 1) = \frac{1}{3} \left[ (4 - 4 a + 2 a^2) m_2(t) + 2(2-a) a m_1^2\right] 
\mathrm{,}
\label{eq:itermomentm}
\end{equation}
As can be easily checked, this reduces to the formula valid for the Z-model, Equation (\ref{eq:momenteq2}), in the case $a = 1$. However a very interesting difference appears in this iteration law (again, an iteration derived from the iteration for the probability distribution with no other assumption than the existence of the second moment). Actually this iteration may lead to an exponentially growing second moment. This happens if the coefficient of $m_2(t) $ in Equation (\ref{eq:itermomentm}) is larger than one. This happens if $a$ is outside of the interval $[1 -\frac{1}{\sqrt{2}}, 1 + \frac{1}{\sqrt{2}}]$ which is compatible with the condition that $0<a<2$. Therefore there can be an instability of the second moment leading to an indefinite increase of the width of the distribution of wealth. Without overstating this, one can say that this makes a model of ever increasing inequality as predicted by some socio-economical theories. 

Moreover, for any $a$ different of $1$, the iteration of higher momenta become unstable. To show this, let us define $b = 1 -a$. The iteration of the $k$-th moment reads:
\begin{equation} 
  m_{k} (t + 1) = \frac{1}{k + 1} \left[ \left((1-b)^k + (1+b)^k\right) m_k(t) + l.o.t(t) \right]  \mathrm{,}
  \label{iter-km}
  \end{equation} 
In this equation, $l.o.t(t)$ is for the lowest order terms, depending on momenta of order less than $k$. Let us consider the smallest  $k$ such that, for a given $a$, there is an exponential growth of this moment.  Therefore $l.o.t(t)$ remains bounded as a function of time and so, if there is an instability, it is dominated after a sufficient number of iterations by the exponentially growing $ \left((1-b)^k + (1+b)^k\right) m_k(t) $. A little algebra shows that the coefficient of $m_k(t)$ on the right-hand side of Equation (\ref{iter-km}) is larger than 1 and the moment grows exponentially if $$\ln(1 + |b|) > \frac{\ln(k +1)}{k}  \mathrm{,}$$
If $|b|$ is small, this is equivalent to the condition 
$$ k > \frac{\ln(1/|b| + 1)}{|b|}  \mathrm{.}$$
It shows that, however $|b|$ is small but not zero, the large order momenta are unstable under the iteration. Recall that $|b|$ small is equivalent to have a mZ-model formally close to the original Z-model. This also shows that, however small (but non zero) $|b|$ is, the steady distribution, if it exists, given by the iteration law should decay with a power law at large values of its argument to make diverge momenta with a large power. It is planned to return to this mathematically interesting question in a future publication.

\section{Diverging moments at time zero}
\label{sec:divmom}

In this Section we return to the Z-model in its original form and consider the following question: what happens to the iterations if the initial momenta diverge beyond a certain power? Indeed, because the initial condition is in principle rather free, provided $m_0 = 1$ and $m_1$ converges, one can always imagine an initial condition with a distribution of wealth decreasing algebraically for large powers. In this case momenta do not exist beyond a certain power. We consider below what happens in this case. In particular we show that, after a finite number of iterations, one recovers a converging moment with a power less than a value increasing as the iterations go. 

We shall limit ourselves to situations where $p_0(u)$, the initial distribution of wealth, behaves at large $u$ as a power law, like 
\begin{equation} 
p_0(u) \approx l_{\alpha_0} u^{-\alpha_0}
\mathrm{,}
\label{eq;as1}
\end{equation}
where $ l_{\alpha_0}$ is a positive constant and $\alpha_0$ a positive exponent. To have finite probability and  first momentum (finite total wealth) on must have $\alpha_0>2$. By putting this power law in the right-hand side of the functional iteration (\ref{eq:iter}), one obtains that at time $t = 1$, the distribution of wealth $p_1(u)$ decays with the power law:

\begin{equation} 
p_1(u) \approx l_{\alpha_1} u^{-\alpha_1}
\mathrm{,}
\label{eq;as2}
\end{equation}
where $\alpha_1 = 2 \alpha_0 - 1$ and where 
$$ l_{\alpha_1} = l_{\alpha_0}^2 B(\alpha_0) \mathrm{,}$$
where 
$$ B(\alpha) =  \int \int_{S(1)} {\mathrm{d}}u' {\mathrm{d}}v' \frac{(u'v')^{-\alpha}}{u' + v'} $$ is a numerical function of the argument $\alpha$. As the iteration formula shows, $\alpha$ increases as the iteration goes and so, as soon as it becomes big enough, momenta of a given power begin to exist, and follow later the explicit recursion formulae given in Equation (\ref{eq:momenteq}). This is correct because momenta of higher and higher order begin to converge the later as their power increases. Therefore the right-hand side of the recursion equation becomes all well defined when the highest moment becomes well defined, all momenta of a smaller power being already finite at this time.

\section{Probability distribution of the wealth of the richest man}
\label{richest}

Looking at the economic magazines, one is struck by their insistence on various lists of rich, if not very rich people, lists ordered according to their supposed wealth. Therefore it is of some interest to consider the question of the distribution of biggest wealth that can be reached within the models outlined in this work.  We begin with a basic question of probability: given a probability distribution $p(x)$, and a number $\nu$ of independent trials, what is the largest value reached among those trials? This  interesting question can be answered quite simply as demonstrated below. Then we apply this result to the case of the Z and of the mZ model. 

Consider first the following problem: given $x_0$ positive, let us draw a number $x$ with probability distribution $p(x)$. What is the distribution of the maximum of $x_0$ and $x$, a maximum denoted as $X$? If $x$ is less than $x_0$ this maximum is $x_0$, in the opposite case it is $x$. Define $$ N(x) = \int_0^x {\mathrm{d}}x' p(x') \mathrm{.}$$ The probability that $x$ is less than $x_0$ is $N(x_0)$. Therefore the probability distribution of $X$ is 
\begin{equation}
\Pi(X, x_0) = N(X) \delta(X-x_0) + p(X) H(X - x_0)
\mathrm{,}
\label{eq:prob1}
\end{equation}
where $H(.)$ is Heaviside function equal to 1 if its argument is positive and zero otherwise. The probability distribution $\Pi(X, x_0)$ is normalised in such a way that 
$$ \int_0^{\infty}  {\mathrm{d}}X \Pi(X, x_0) = 1 \mathrm{,}$$ a consequence of the property $N(\infty) = 1$. 

\begin{figure}[t]
\begin{center}
\includegraphics[scale=0.5]{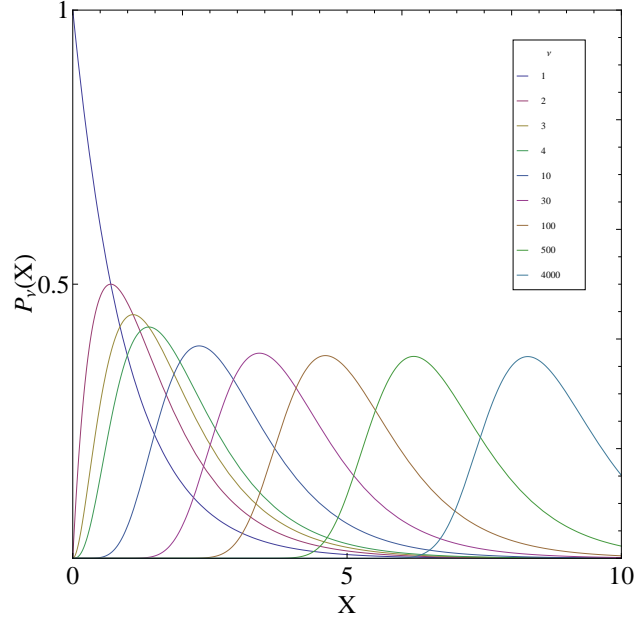}
\end{center}
\caption{$P_{\nu}(X)$ for different $\nu$ when $p(x)=e^{-x}$. Observe the monotonic increasing of $X_{\nu}$ with $\nu$. For this case, when $\nu\gg 10$, observe that $P_{\nu}(X_{\nu})$ is constant and $P_{\nu}(X)$ presents a soliton-like waveform.}
\label{fig1}
\end{figure}

\begin{figure}[t]
\begin{center}
\includegraphics[scale=0.5]{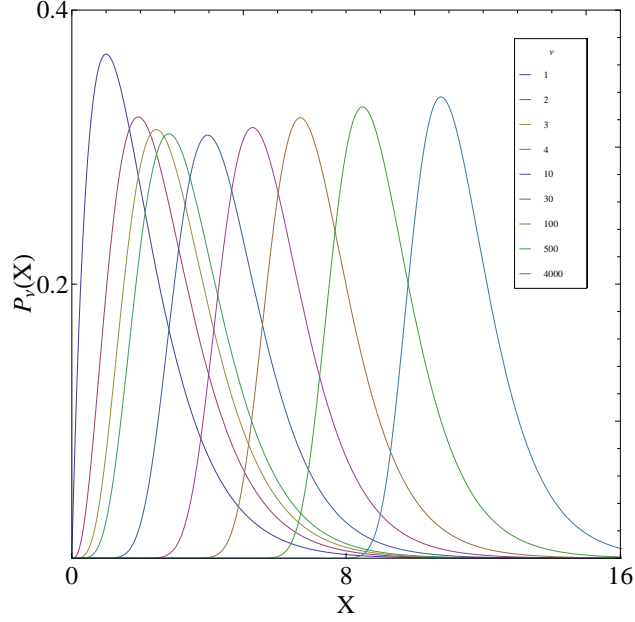}
\end{center}
\caption{$P_{\nu}(X)$ for different $\nu$ when $p(x)=xe^{-x}$. Observe the monotonic increasing of $X_{\nu}$ with $\nu$. For this case, when $\nu\gg 10$, observe that $P_{\nu}(X_{\nu})$ is not constant and $P_{\nu}(X)$ presents an increase of the maximal wave amplitude.}
\label{fig2}
\end{figure}

Suppose now that $x_0$, instead of being taken as a fixed number is drawn at random with a probability distribution $q(x_0)$. Therefore the probability distribution of the maximum of $x$ and $x_0$ has to be averaged over the choices of $x_0$. This yields 
\begin{equation}
P(X) = \int_0^{\infty}  {\mathrm{d}}x_0 q(x_0) \Pi(X, x_0)  = N(X) q(X) + p(X)\int_0^X  {\mathrm{d}}x' q(x')
\mathrm{,}
\label{eq:prob2}
\end{equation}
One can check by performing the integrals in the quadrant $x, x' >0$ that $$\int_0^{\infty}  {\mathrm{d}}X P(X) = \int_0^{\infty}  {\mathrm{d}}x p(x) \int_0^{\infty}  {\mathrm{d}}x' q(x') = 1\mathrm{.}$$
 From equation (\ref{eq:prob2}) one can derive the probability distribution of the largest value drawn after $\nu$ (integer) independent trials , each one with the probability distribution $p(x)$. Let $P_{\nu}(x)$ be the probability distribution of the maximum of $\nu$ trials. After one trial $P_{1}(X) = p(X)$. From equation (\ref{eq:prob2}) one derives the recursion formula between $P_{\nu}(X)$ and $P_{\nu+1}(X)$:
\begin{equation}
P_{\nu+1}(X) = N(X) P_{\nu}(X) + p(X)\int_0^X  {\mathrm{d}}x' P_{\nu}(x')
\mathrm{,}
\label{eq:prob3}
\end{equation}
Define now $Q_{\nu}(X) = \int_0^X  {\mathrm{d}}x' P_{\nu}(x')$. This allows to write equation (\ref{eq:prob3}) like:
\begin{equation}
\frac{{\mathrm{d}}Q_{\nu+1}(X)}{{\mathrm{d}}X}= N(X) \frac{{\mathrm{d}}Q_{\nu}(X)}{{\mathrm{d}}X} + \frac{{\mathrm{d}}N(X)}{{\mathrm{d}}X} Q_{\nu}(X)
\mathrm{,}
\label{eq:prob4}
\end{equation}
This can be obviously integrated as 
$$Q_{\nu+1}(X) = N(X) Q_{\nu}(X) + S_{\nu} \mathrm{,}$$ 
where $S_{\nu}$ is a constant of integration, independent on $X$. Because $Q_{\nu}(0) = 0 $ for all $S_{\nu}$, $S_{\nu} = 0$ also for all ${\nu}$. Therefore 
\begin{equation}
Q_{\nu} (X)= \left (\int_0^X  {\mathrm{d}}x' p(x') \right)^{\nu} 
\mathrm{,}
\label{eq:prob5}
\end{equation}
and 
\begin{equation}
P_{\nu} (X)= \nu p(X) \left (\int_0^X  {\mathrm{d}}x' p(x') \right)^{\nu -1} 
\mathrm{.}
\label{eq:prob6}
\end{equation}
Suppose $p(x)$ is a smooth function decaying continuously to zero as $x$ tends to infinity.  In this case it is possible to get the asymptotic form of $P_{\nu} (X)$ at $\nu$ very large. Let us write $P_{\nu} (X)$ as an exponential 
$$ P_{\nu} (X)= e^{T(\nu, X)} \mathrm{.}$$
with 
$$ T(\nu, X) =  \ln(\nu) + \ln(p(X)) + (\nu -1)  \ln\left(\int_0^X  {\mathrm{d}}x' p(x') \right) \mathrm{.}$$
In the limit $\nu$ large, one expects that the distribution $ P_{\nu} (X)$ has more and more weight at larger and larger values of $X$, which is also what is found by looking numerically at the shape of $ P_{\nu} (X)$  in this limit for various possible $p(X)$. See Figs. \ref{fig1} and \ref{fig2}. Therefore, in this limit, $P_{\nu} (X)$ should become more and more concentrated around the value of $X$ such that the derivative  $\frac{\partial T(\nu, X)}{\partial X} = 0$. This derivative vanishes when $X$ is the root $X_{\nu}$ of 
$$ \nu = 1 - \frac{p'\cdot N}{p^2}  \mathrm{,}$$ where $p' = \frac{dp}{dX}$. 
When $X_{\nu}$ is large, then $N(X_{\nu})=\int_0^{X_{\nu}} {\mathrm{d}}x' p(x') \approx 1$.
At $\nu$ large, this root $X_{\nu}$ is unique and large. This can be seen by noticing that $ - \frac{p'}{p^2}  =  \frac{d(1/p)}{dX} $, and by assuming that $1/p$ is a smooth function increasing monotonically to infinity as $x$ tends to infinity. To make its first momentum $m_1$ convergent $p(X)$ must decay faster than $x^{-2}$ at infinity, so that the derivative $ \frac{d(1/p)}{dX} $ must grow faster than $X$ at $X$ large. Therefore the function  $X_{\nu}$ grows slower than $\nu$ as $\nu$ tends to infinity but it grows to infinity for any function $p(x)$ tending smoothly to zero as $x$ tends to infinity. This growth will depend on the behaviour of $p(x)$ as $x$ tends to infinity.

The function $X_{\nu}$ gives the order of magnitude of the maximum wealth after $\nu$ iterations. By continuing the expansion of $ T(\nu, x)$ near $X_{\nu}$ to the quadratic order with respect to the difference $\delta X = X - X_{\nu}$, 
one finds that 
$$ T(\nu, X)  \approx T(\nu, X_{\nu}) + \frac{\delta X^2}{2} \frac{\partial^2  T(\nu, X)}{\partial X^2} +...\mathrm{.}$$
where the second derivative is computed at $X = X_{\nu}$.

After some algebra and by taking into account that $\nu$ is large and that, in the limit $X_{\nu}$ large, $N(X_{\nu}) \approx 1$, one finds 
$$ \frac{\partial^2  T(\nu, X)}{\partial X^2} \approx - p( X_{\nu}) \frac{d^2 (1/p)}{dX^2}\mathrm{.}$$
To prove that the width of the maximum of the distribution is much less than $X_{\nu}$, one can do the following approximate scaling argument. We have $\frac{p'}{p^2} \approx - \nu$. Assuming that $p' \approx \frac{p(X_{\nu})}{X_{\nu}}$, which is certainly correct for a probability distribution $p(.)$ decaying like a power law at large arguments, one finds $\nu \sim \frac{1}{ X_{\nu} p( X_{\nu})}$. Using the same kind of scaling argument one finds that 
$$ \frac{\partial^2  T(\nu, X)}{\partial X^2} \sim \frac{-\nu  p(X_{\nu})}{X_{\nu}} \sim -\frac{1}{X_{\nu}^2} \mathrm{.}$$
This shows that, at least for distributions $p(x)$ decaying like power laws, the width of the probability distribution $ P_{\nu} (X)$ is of order $X_{\nu}$ for $\nu$ very large, although its center is at $X_{\nu}$. In this case the width of the probability distribution and its center are large and of the same order of magnitude. Therefore one may guess that it behaves like 
$$ P_{\nu} (X) \approx \frac{1}{X_{\nu}} \hat{P}\left(\frac{X}{X_{\nu}}\right) \mathrm{.}$$ where $\hat{P}$ is a positive numerical function of order one when its argument is of order one. It is normalized in such a way that $\int_0^{\infty} 
\hat{P}(z) {\mathrm{d}} z = 1$. From the derivation,  this function depends on the way $p(x)$ behaves as $x$ tends to infinity.

\section{Conclusions and perspectives}
\label{sec:concper}

 Thanks to its mathematical structure the Z-model can be solved and somehow extended to bring interesting results with, perhaps, a connection to the complicated phenomenology of real economics. Despite its strongly nonlinear character it can be solved without assuming too many things. A remarkable feature of this model is its convergence to an exponential distribution of wealth. Of course any difference between reality and this model may have many explanations. Among others, it has been suggested, such as one of us (YP) also suggested it during the Noma-13 conference, that this model lacks an important element present in economies of developed countries, the tax system, with a more or less explicit claim of redistributing the wealth.  Such a tax system could be perhaps represented by adding a third partner in each binary transaction, taking its pound of flesh at the transaction and redistributing it randomly at the next step, more or less the way the VAT tax (added value tax) works. This paper introduces also a modified Z-model, where at each transaction money is exchanged which is not actually possessed by the economic agents, something occuring all the time in modern economies. Amazingly this induces an instability in the distribution of wealth and makes grow indefinitely the higher momenta of its distribution, even though the total amount remains the same. Although this happens in a very idealised model, it could be closer to reality than the original Z-model with its rather narrow distribution of wealth.  
  
\date{\today }

\end{document}